\documentclass[11pt,a4paper]{article}
\usepackage{amssymb,amsmath,a4wide,graphicx}
\newcommand{\avg}[1]{\langle #1\rangle}
\newcommand{\req}[1]{(\ref{#1})}
\newcommand{\K}{_{\mathrm{K}}}

\newcommand{\dd}{\mathrm{d}}
\newcommand{\cR}{\mathcal{R}}
\begin{document}
\title{Diversification and limited information
in the Kelly game}
\author{Mat\'u\v s Medo$^1$, Yury M. Pis'mak$^2$ and
Yi-Cheng Zhang$^{1,3}$}
\date{}
\maketitle

{\small\noindent
$^1$ Physics Department, University of Fribourg, Chemin du
Mus\'ee 3, 1700 Fribourg, Switzerland\\
$^2$ Department of Theoretical Physics, State University of
Saint-Petersburg, 198 504 Saint-Petersburg, Russian Federation\\
$^3$ Lab of Information Economy and Internet Research,
University of Electronic Science and Technology, 610054 Chengdu,
China}

\begin{quote}
Financial markets, with their vast range of different investment
opportunities, can be seen as a system of many different
simultaneous games with diverse and often unknown levels of risk
and reward.  We introduce generalizations to the classic Kelly
investment game [Kelly (1956)] that incorporates these features,
and use them to investigate the influence of diversification and
limited information on Kelly-optimal portfolios.  In particular
we present approximate formulas for optimizing diversified
portfolios and exact results for optimal investment in unknown
games where the only available information is past outcomes.
\end{quote}

\section{Introduction}
Portfolio optimization is one of the key topics in finance. It
can be characterized as a~search for a satisfactory compromise
between maximization of the investor's capital and minimization
of the related risk. The outcome depends on properties of the
investment opportunities and on the investor's attitude to risk
but crucial is the choice of the optimization goals. In last
decades, several approaches to portfolio optimization have been
proposed---good recent overviews of the field can be found
in~\cite{Duff01,EGBG06}.

In this paper we focus on the optimization strategy proposed by
Kelly~\cite{Ke56} where repeated investment for a long run is
explored. As an optimization criterion, maximization of the
average exponential growth rate of the investment is suggested.
This approach has been investigated in detail in many subsequent
works~\cite{Ma76,FiWh81,RoTh92,MZ98,Br00,our} and it is optimal
according to various criteria~\cite{Br61,Th00}. Similar ideas
lead to the universal portfolios described in~\cite{Co91}.

While the original concept focuses on a single investment in
many successive time periods, we generalize it to a diversified
investment. This extension is well suited for investigating the
effects of diversification and limited information on investment
performance. However, in complex models of real investments,
important features can get unnoticed. Therefore we replace
realistic assumptions about the available investment
opportunities (\emph{e.g.} log-normal distribution of returns)
by simple risky games with binary outcomes. While elementary,
this setting allows us to model and analytically treat many
investment phenomena; all scenarios proposed and investigated
here are meant as metaphors of real-life problems.

The paper is organized as follows. In section \ref{sec-summary}
we briefly overview the original Kelly problem and the main
related results. In section \ref{sec-simult} we allow investment
in simultaneous risky games and investigate the resulting
portfolio diversification. In section \ref{sec-outsider} it is
shown that an investment profiting from additional information
about one game (an insider approach) can be outperformed by
a diversified investment (an outsider approach). Finally in
section \ref{sec-memory} we investigate the case where
properties of a risky game are unknown and have to be inferred
from its past outcomes. We show that in consequence, the Kelly
strategy may be inapplicable.

\section{Short summary of the Kelly game}
\label{sec-summary}
In the original Kelly game, an investor (strictly speaking,
a~gambler) with the starting wealth $W_0$ is allowed to
repeatedly invest a~part of the available wealth in a risky
game. In each turn, the risky game has two possible outcomes:
with the probability $p$ the stake is doubled, with the
complementary probability $1-p$ the stake is lost. It is assumed
that the winning probability $p$ is constant and known to the
investor. We introduce the game return $R$ which is defined as
$R:=(W_r-W_i)/W_i$ where $W_i$ is the invested wealth and $W_r$
is the resulting wealth. For the risky game described above the
possible returns per turn are $+1$ (win results in $W_r=2W_i$)
and $-1$ (loss results in $W_r=0$). Investor's consumption is
neglected.

Since properties of the risky game do not change in time, the
investor bets the same fraction $f$ of the actual wealth in each
turn. The investor's wealth follows a multiplicative process and
after $N$ turns it is equal to
\begin{equation}
\label{WN}
W_N(R_1,\dots,R_N)=W_0\prod_{i=1}^N (1+fR_i)
\end{equation}
where $R_i$ is the game return in turn $i$. Since the successive
returns $R_i$ are independent, from Eq.~\req{WN} the average
wealth after $N$ turns can be written as (averages over
realizations of the risky game we label as $\avg{\cdot}$)
\begin{equation}
\avg{W_N}=W_0\,\avg{1+fR_i}^N=W_0\,\big[1+(2p-1)f\big]^N.
\end{equation}
Maximization of $\avg{W_N}$ can be used to optimize the
investment. Since for $p<1/2$, $\avg{W_N}$ is a decreasing
function of $f$, the optimal strategy is to refrain from
investing, $f^*=0$. By contrast, for $p>1/2$ the quantity
$\avg{W_N}$ increases with $f$ and thus the optimal strategy is
to stake everything in each turn, $f^*=1$. Then, while
$\avg{W_N}$ is maximized, the probability of getting ruined in
first $N$ turns is $1-p^N$. Thus in the limit $N\to\infty$, the
investor bankrupts inevitably and maximization of $\avg{W_N}$ is
not a good criterion for a long run investment.

In his seminal paper~\cite{Ke56}, Kelly suggested maximization
of the exponential growth rate of the investor's wealth
\begin{equation}
\label{G}
G=\lim_{N\to\infty}\frac1N\log_2\frac{W_N}{W_0}
\end{equation}
as a criterion for investment optimization (without affecting
the results, in our analysis we use natural logarithms). Due to
the multiplicative character of $W_N$, $G$ can be rearranged as
\begin{equation}
G=\lim_{N\to\infty}\frac1N\sum_{i=1}^N\ln\big(1+fR_i\big)=
\avg{\ln W_1},
\end{equation}
Notice that while we investigate repeated investments, wealth
$W_1$ after turn step plays a prominent role in the
optimization. For the risky game introduced above is
$\avg{\ln W_1}=p\ln(1+f)+(1-p)\ln(1-f)$ which is maximized by
the investment fraction
\begin{equation}
\label{f-Kelly}
f\K(p)=2p-1.
\end{equation}
When $p<1/2$, $f\K<0$ (a short position) is suggested. In this
paper we exclude short-selling and thus for $p<1/2$ the optimal
choice is $f\K=0$. For $p\in[1/2,1]$, the maximum of $G$ can be
rewritten as
\begin{equation}
\label{G-Kelly}
G\K(p)=\ln 2-S(p),
\end{equation}
where $S(p)=-\big[p\ln p+(1-p)\ln(1-p)\big]$ is the entropy
assigned to the risky game with the winning probability $p$.

There is a parallel way to Eq.~\req{f-Kelly}. If we define the
compounded return per turn $R_N$ by the formula
$W_N=W_0\,(1+R_N)^N$ and its limiting value by
$\cR:=\lim_{N\to\infty}R_N$, it can be shown that
$\cR=\exp[G]-1$. Thus maximization of $\cR$ leads again to
Eq.~\req{f-Kelly}. Quoting Markowitz in~\cite{Ma76}, Kelly's
approach can be summarized as ``In the long-run, thus defined,
a~penny invested at $6.01\%$ is better---eventually becomes and
stays greater---than a million dollars invested at $6\%$.'' While
$G$ is usually easier to compute than $\cR$, in our discussions
we often use $\cR$ because it is more illustrative in the context
of finance. Using $\cR\K=\exp[G\K]-1$, the maximum of $\cR$ can
be written as
\begin{equation}
\label{cR-Kelly}
\cR\K(p)=2p^p(1-p)^{1-p}-1.
\end{equation}
When $p=1/2$, $\cR\K=0$; when $p\to1$, $\cR\K=1$.

The results obtained above we illustrate on a particular risky
game with the winning probability $p=0.6$. Since $p>0.5$, it is
a profitable game and a gambler investing all the available
wealth has the expected return $\avg{R}=2p-1=20\%$ in one turn.
However, according to Eq.~\req{f-Kelly} in the long run the
optimal investment fraction is $f\K=0.2$. Thus, the expected
return in one turn is reduced to $0.2\times 20\%=4\%$. For
repeated investment, the average compounded return $\cR$
measures the investment performance better. From
Eq.~\req{cR-Kelly} it follows that for $p=0.6$ is $\cR=2.0\%$.
We see that a wise investor gets in the long run much less than
the illusive return $20\%$ of the given game (and a~naive
investor gets even less). In the following section we
investigate how diversification (if possible) can improve this
performance.

\section{Simultaneous independent risky games}
\label{sec-simult}
We generalize the original Kelly game assuming that there are
$M$ independent risky games which can be played simultaneously
in each time step (correlated games will be investigated in a
separate work). In game $i$ ($i=1,\dots,M$) the gambler invests
the fraction $f_i$ of the actual wealth. Assuming fixed
properties of the games, this investment fraction again does not
change in time. For simplicity we assume that all games are
identical, \emph{i.e.} with the probability $p$ is $R_i=1$ and
with the probability $1-p$ is $R_i=-1$. Consequently, the
optimal fractions are also identical and the investment
optimization is simplified to a~one-variable problem where
$f_i=f$.

For a given set of risky games, there is the probability
$(1-p)^M$ that in one turn all $M$ games are loosing. In
consequence, for all $p<1$ the optimal investment fraction $f^*$
is smaller than $1/M$ and thus $Mf^*<1$ (otherwise the gambler
risks getting bankrupted and the chance that this happens
approaches one in the long run). If in one turn there are $w$
winning and $M-w$ loosing games, the investment return is
$(2w-M)f$ and the investor's wealth is multiplied by the factor
$1+(2w-M)f$. Consequently, the exponential growth rate is
\begin{equation}
\label{G-simult}
G=\avg{\ln W_1}=\sum_{w=0}^M P(w;M,p)\ln\big[1+(2w-M)f\big],
\end{equation}
where $P(w;M,p)=\binom{M}{w}p^w(1-p)^{M-w}$ is a binomial
distribution. The optimal investment fraction is obtained by
solving $\partial G/\partial f=0$. If we rewrite
$2w-M=[f(2w-M)+1-1]/f$ and use the normalization of $P(w;M,p)$,
we simplify the resulting equation to
\begin{equation}
\label{f-condition}
\sum_{w=0}^M\frac{P(w;M,p)}{1+(2w-M)f}=1.
\end{equation}
For $M=1$ we obtain the well-known result $f_1^*=2p-1$, for
$M=2$ the result is $f_2^*=(2p-1)/(4p^2-4p+2)$. Formulae for
$M=3,4$ are also available but too complicated to present here.
For $M\geq5$, Eq.~\req{f-condition} has no closed solution and
thus in the following sections we seek for approximations. In
complicated cases where such approximations perform badly,
numerical algorithms are still applicable~\cite{Wh07}.

\subsection{Approximate solution for an unsaturated portfolio}
\label{ssec-weak}
By an unsaturated portfolio we mean the case when a small part
of the available wealth is invested, $Mf^*\ll1$. Then also
$\vert(2w-M)f^*\vert\ll1$ and in Eq.~\req{f-condition} we can
use the expansion $1/(1+x)\approx1-x+x^2\pm\dots$ ($|x|<1$).
Taking only the first three terms into account, we obtain
$\sum_{w=0}^M P(w;M,p)[1-f(2w-M)+f^2(2w-M)^2]=1$ and after the
summation we get
\begin{equation}
\label{f-small}
f^*(p)=\frac{2p-1}{M(2p-1)^2+4p(1-p)}.
\end{equation}
When $p-1/2\ll1/M$, $f^*(p)$ simplifies to $f^*=2p-1$, the
gambler invests in each game as if other games were not present.
When the available games are diverse, this result generalizes to
$f_i^*=2p_i-1$. For $M=1,2$, Eq.~\req{f-small} is equal to the
exact results obtained above.

\subsection{Approximate solution for a saturated portfolio}
\label{ssec-strong}
By a saturated portfolio we mean the case when almost all
available wealth is invested, $1-Mf^*\ll1$. The extreme is
achieved for $p=1$ when all wealth is distributed evenly among
the games. We introduce the new variable $x:=1/M-f$ and rewrite
Eq.~\req{f-condition} as
$$
\frac{P(0;M,p)}{xM}+
\sum_{w=1}^M \frac{P(w;M,p)}{2w/M-x(2w-M)}=1.
$$
Since according to our assumptions $0<x\ll1/M$, to obtain the
leading order approximation for $f^*$ we neglect $x$ in the sum
which is then equal to $\tfrac{M}2\avg{1/w}$. The crude
approximation $\avg{1/w}\approx1/\avg{w}$ leads to the result
\begin{equation}
\label{f-big}
f^*=\frac1M\bigg[1-\frac{2p(1-p)^M}{2p-1}\bigg].
\end{equation}
As expected, in the limit $p\to1$ we obtain $f^*=1/M$. When the
available games are diverse, this approximation does not work
well and in the optimal portfolio, the most profitable games
prevail.

\begin{figure}
\centering
\includegraphics[scale=0.26]{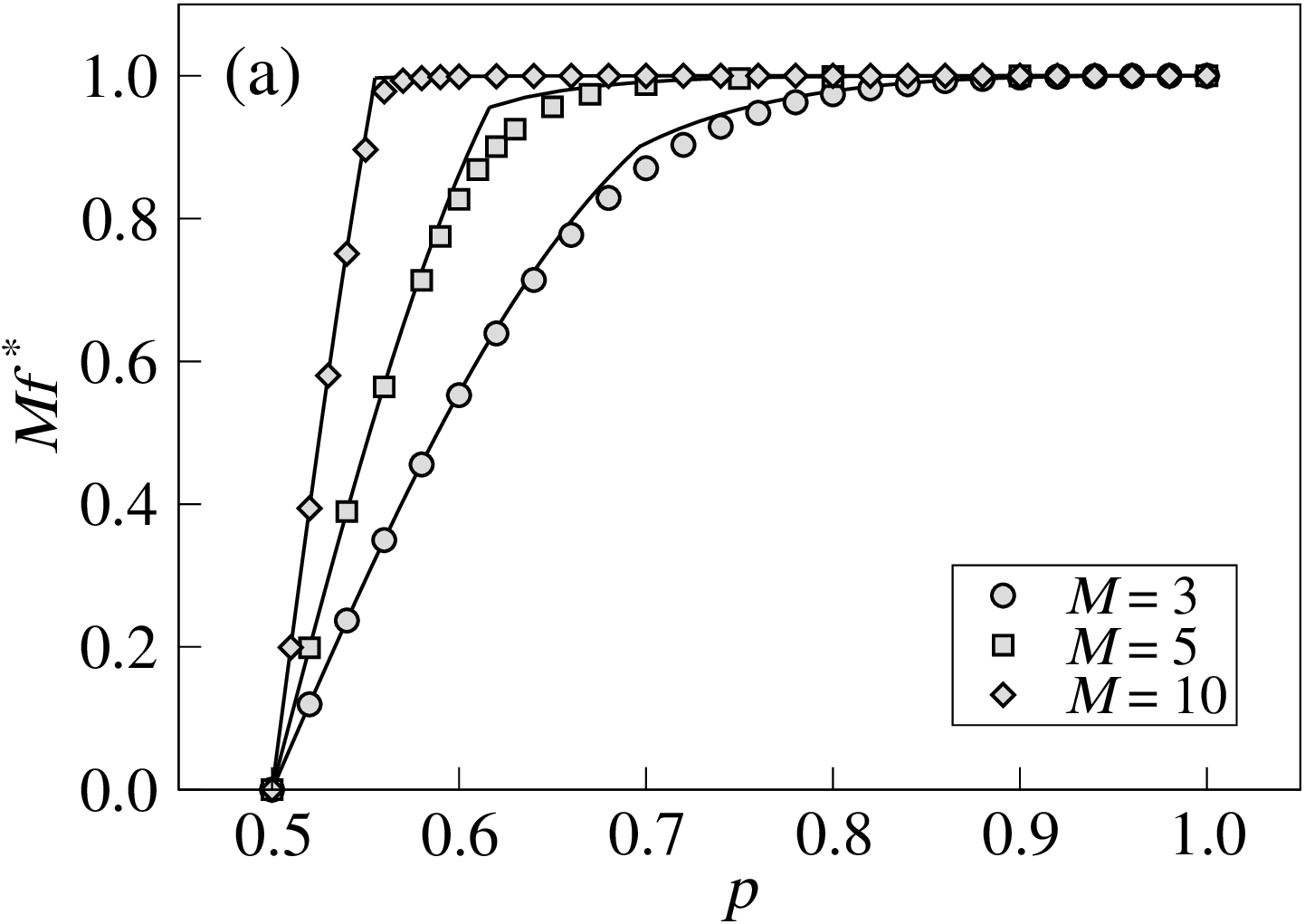}\quad
\includegraphics[scale=0.26]{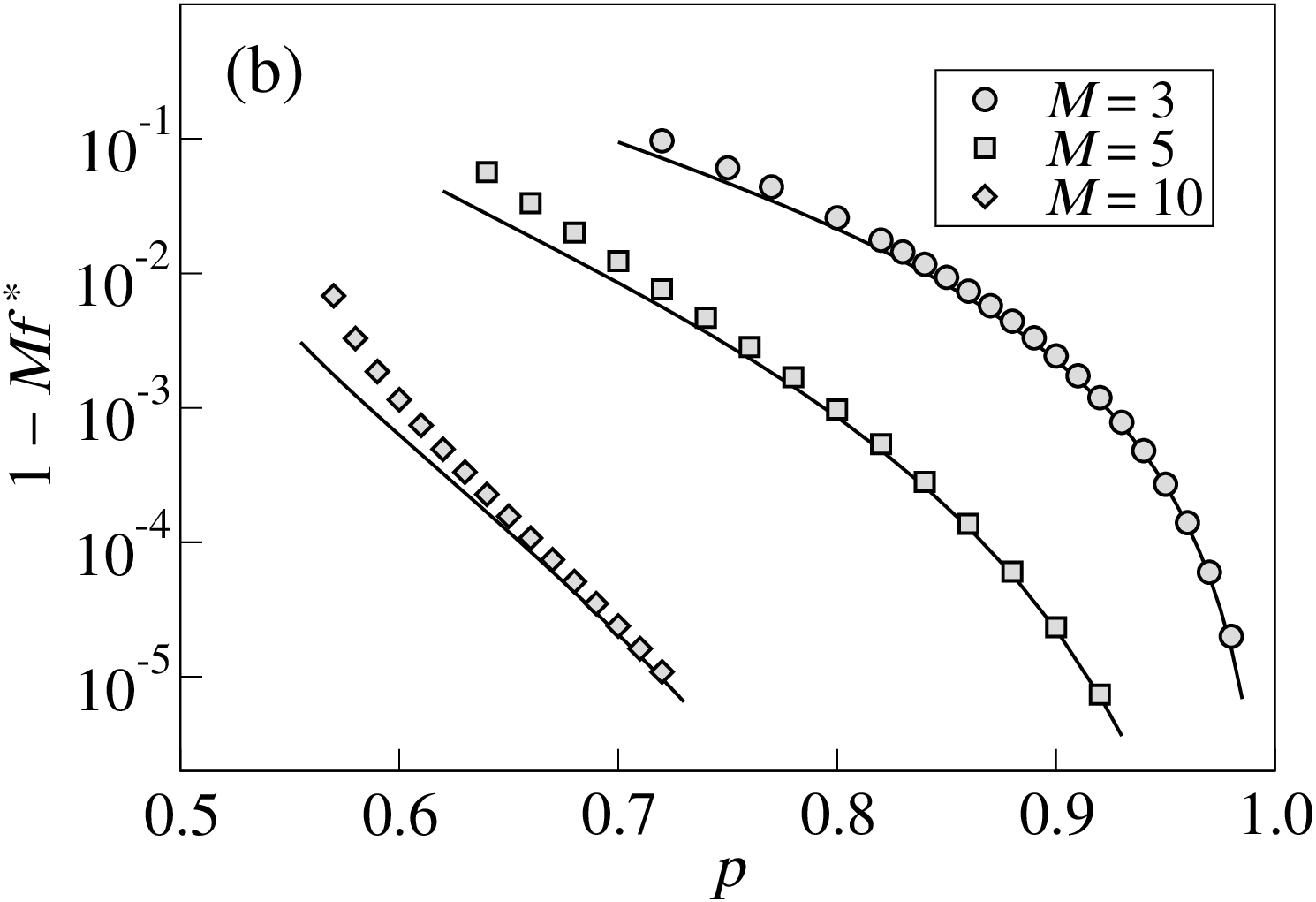}
\caption{The comparison of numerical results for the optimal
investment fraction $f^*$ (obtained using Mathematica, shown as
symbols) with the analytical results given in Eq.~\req{f-small}
and Eq.~\req{f-big} (shown as solid lines). (a) The total
investment $Mf^*$ as a function of $p$. (b) To judge better the
approximation for a saturated portfolio, the univested fraction
$1-Mf^*$ is shown as a function of $p$.}
\label{fig-Mf}
\end{figure}

Approximations Eq.~\req{f-small} and Eq.~\req{f-big} can be
continuously joined if for $p\in[\tfrac12,p_c]$ the first one
and for $p\in(p_c,1]$ the second one is used; the boundary value
$p_c$ is determined by the intersection of these two results.
A~comparison of the derived approximate results with numerical
solutions of Eq.~\req{f-condition} is shown in
Fig.~\ref{fig-Mf}. For most parameter values a good agreement
can be seen, the largest deviations appear for a mediocre number
of games ($M\simeq5$) and a mediocre winning probability
($p\simeq p_c$).

\section{Diversification vs information}
\label{sec-outsider}
In real life, investors have only limited information about the
winning probabilities of the available risky games. These
probabilities can be inferred using historical wins/losses data
but these results are noisy and the analysis requires investor's
time and resources (the process of inference is investigated in
detail in Sec.~\ref{sec-memory}). At the same time, insider
information can improve the investment performance
substantially. A similar insider-outsider approach is discussed
in the classical paper on efficient markets~\cite{GS80} and in
a~simple trading model~\cite{CZ00}. We model the described
situation by a competition of two investors who can invest in
multiple risky games; each of the games has the winning
probability alternating with even odds between $p+\Delta$ and
$p-\Delta$ ($1/2<p\leq1$, $0\leq\Delta\leq 1-p$). The insider
focuses on one game in order to obtain better information about
it---we assume that the exact winning probability is available
to him. By contrast, the outsider invests in several games but
knows only the time average $p$ of the winning probability. We
shall investigate when the outsider performs better than the
insider.

The insider knows the winning probability and thus can invest
according to Eq.~\req{f-Kelly}. If $p-\Delta>1/2$, he invests in
each turn, if $p-\Delta\leq1/2$, he invests only when the
winning probability is $p+\Delta$. Combining the previous
results, the exponential growth rate of the insider
$G_I=\avg{\ln W_1}$ can be simplified to
\begin{equation}
\label{G-insider}
G_I=\begin{cases}
\frac12[\ln 2+S(p+\Delta)]&p-\Delta\leq 1/2,\\
\frac12[\ln 2+S(p+\Delta)]+
\frac12[\ln 2+S(p-\Delta)]&p-\Delta>1/2,
\end{cases}
\end{equation}
where $S(p)$ is the same as in Eq.~\req{G-Kelly}. We assume that
the outsider invests in $M$ identical and independent risky
games. For him, each risky game is described by the average
winning probability $p$. Consequently, the exponential growth
rate of his investment is given by Eq.~\req{G-simult} and for
the optimal investment fractions results from the previous
section apply.

\begin{figure}
\centering
\includegraphics[scale=0.26]{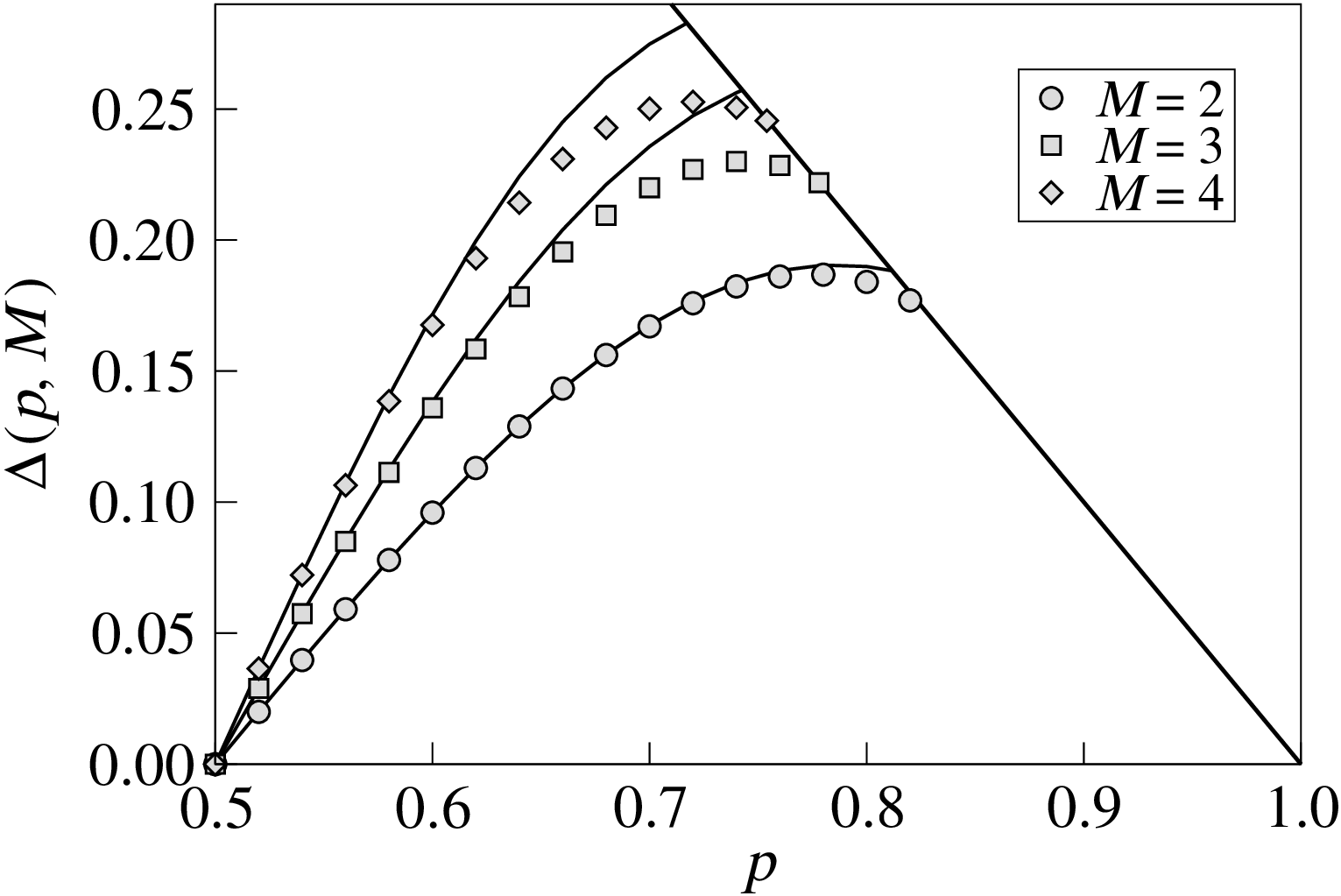}
\caption{The limiting value $\Delta(p,M)$ when the investment
performances of the insider investing in one game and the
outsider investing in $M$ games are equal. Numerical solutions
of Eq.~\req{Delta} are shown as symbols, analytical results
obtained as described in the text are shown as lines.}
\label{fig-phases}
\end{figure}

The limiting value of $\Delta$, above which the insider performs
better that the outsider, is given by
\begin{equation}
\label{Delta}
G_I(p,\Delta)=G_O(p,M).
\end{equation}
Due to the form of $G_I(p,\Delta)$, it is impossible to find an
analytical expression for $\Delta$. An approximate solution can
be obtained by expanding $G_I(p,\Delta)$ in powers of $\Delta$;
first terms of this expansion have the form
$$
G_I(p,\Delta)=\begin{cases}
\tfrac12\,\Big[G\K(p)+\Delta\big(\ln p-\ln[1-p]\big)+
\frac{\Delta^2}2\Big(\frac1p+\frac1{1-p}\Big)\Big]
&p-\Delta\leq 1/2,\\
G\K(p)+\frac{\Delta^2}2\Big(\frac1p+\frac1{1-p}\Big)+
\frac{\Delta^4}{12}\Big(\frac1{p^3}+\frac1{(1-p)^3}\Big)
&p-\Delta>1/2.\\
\end{cases}
$$
By substituting this to Eq.~\req{Delta} we obtain a quadratic
(when $p-\Delta\leq1/2$) or biquadratic (when $p-\Delta>1/2$)
equation for $\Delta$ which can be solved analytically. In this
way we get $\Delta(p,M)$, for $\Delta<\Delta(p,M)$ the outsider
performs better than the insider. When $p-1/2\ll1$, the lowest
order approximation is $\Delta(p,M)=(p-1/2)(\sqrt{2M}-1)$. To
review the accuracy of our calculation, in Fig.~\ref{fig-phases}
analytical results for $M=2,3,4$ are shown together with
numerical solutions of Eq.~\req{Delta} obtained using
Mathematica. In line with expectations, the higher is the
winning probability $p$, the harder it is for the insider to
outperform the outsider.

\section{Finite memory problem}
\label{sec-memory}
As already mentioned, in real life investors do not have
information on the exact value of the winning probability $p$,
it has to be inferred from the available past data. In addition,
since $p$ can vary in time, it may be better to focus on a
recent part of the data and obtain a fresh estimate. To model
the described situation we assume that the investor uses
outcomes from the last $L$ turns for the inference and that the
winning probability $p$ is fixed during this period
(a generalization to variable $p$ will be also discussed). The
impact of uncertainty on the Kelly portfolio is investigated
also in~\cite{BrWh96} where certain prior information and
long-term stationarity of $p$ are assumed.

Let's label the number of winning games in last $L$ turns as
$w$ ($w=0,\dots,L$). The resulting knowledge about $p$ can be
quantified by the Bayes theorem~\cite{SS06} as
\begin{equation}
\label{Bayes}
\varrho(p\vert w,L)=\frac{\pi(p)P(w\vert p,L)}
{\int_0^1\pi(p)P(w\vert L,p)\,\mathrm{d} p}.
\end{equation}
Here $\pi(p)$ is the prior probability distribution of $p$ and
$P(w\vert L,p)$ is the probability distribution of $w$ given the
values $p$ and $L$. Due to mutual independence of consecutive
outcomes, $P(w\vert L,p)$ is a binomial distribution and
$P(w\vert L,p)=\binom{L}{w}p^w(1-p)^{L-w}$. We assume that all
information available to the investor is represented by the
observation of previous outcomes, no additional information
enters the inference. The maximum prior ignorance is represented
by $\pi(p)=1$ for $p\in[0,1]$ (a~uniform prior). Consequently,
Eq.~\req{Bayes} simplifies to
\begin{equation}
\label{Bayes-result}
\varrho(p\vert w,L)=\frac{(L+1)!}{w!(L-w)!}\,p^w(1-p)^{L-w}.
\end{equation}
This is the investor's information about $p$ after observing $w$
wins in the last $L$ turns.

When in the Kelly game instead of the winning probability $p$,
only the probability distribution $\varrho(p)$ is known,
maximization of $G=\avg{\ln W_1}$ results in $f^*=2\avg{p}-1$.
We prove this theorem for a special case of two possible winning
probabilities $p_1$ and $p_2$, $P(p_1)=P_1$, $P(p_2)=P_2$,
$P_1+P_2=1$ (extension to the general case is straightforward).
The exponential growth rate can be now written as
\begin{equation*}
G=(P_1p_1+P_2p_2)\,\ln(1+f)+(1-P_1p_1-P_2p_2)\,\ln(1-f).
\end{equation*}
This is maximized by $f^*=2(P_1p_1+P_2p_2)-1$. Since
$P_1p_1+P_2p_2=\avg{p}$, we have $f^*=2\avg{p}-1$. From
Eq.~\req{Bayes-result} follows $\avg{p}=(w+1)/(L+2)$ and
consequently
\begin{equation}
\label{f-memory}
f^*(w,L)=\frac{2w-L}{L+2}
\end{equation}
for $w\geq L/2$. Since we do not consider the possibility of
short selling in this work, $f^*=0$ for $2w<L$. Even when $w=L$
(all observed game are winning), $f^*<1$. This is a consequence
of the noisy information about $p$.

\begin{figure}
\centering
\includegraphics[scale=0.26]{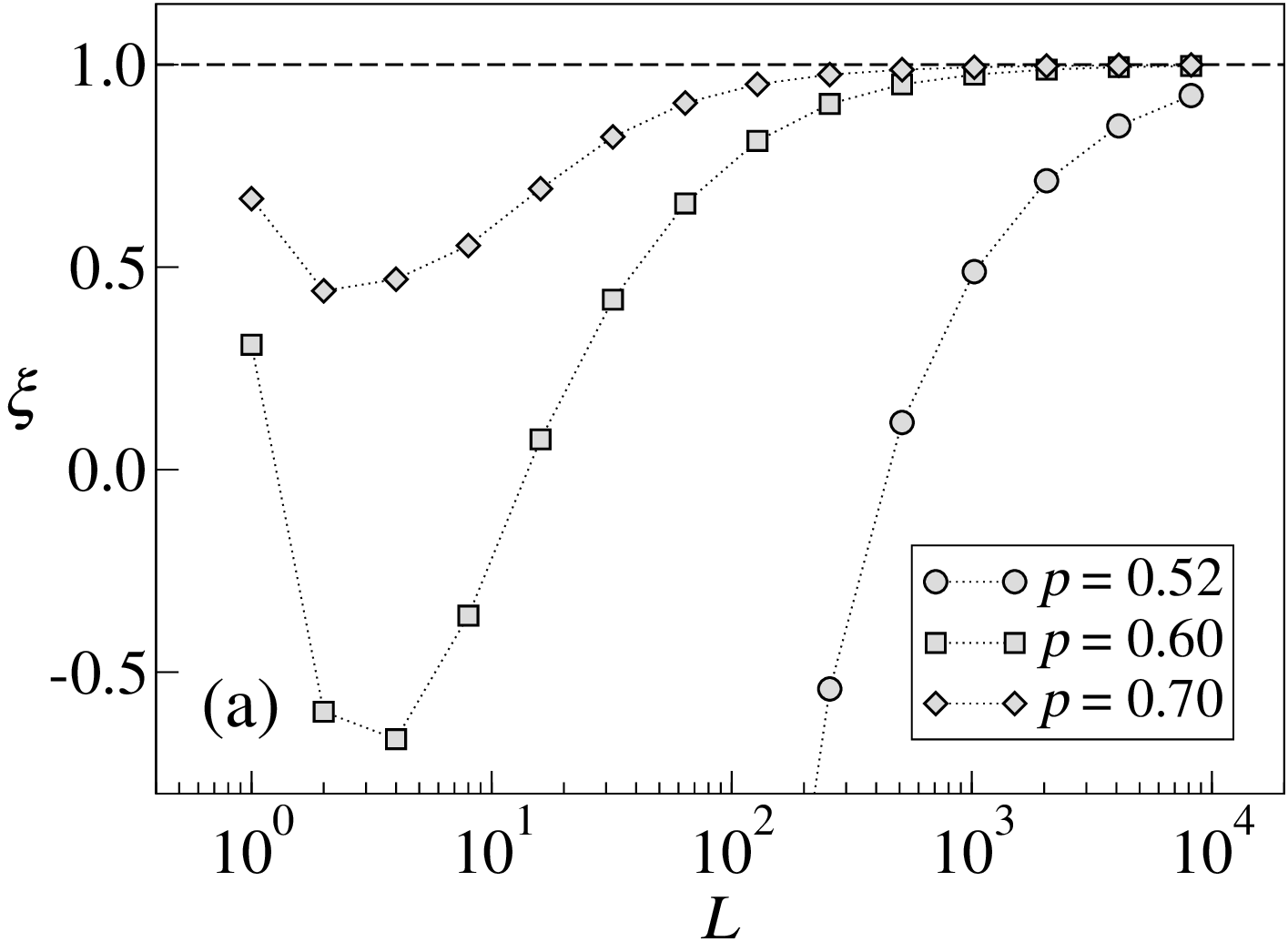}\quad
\includegraphics[scale=0.26]{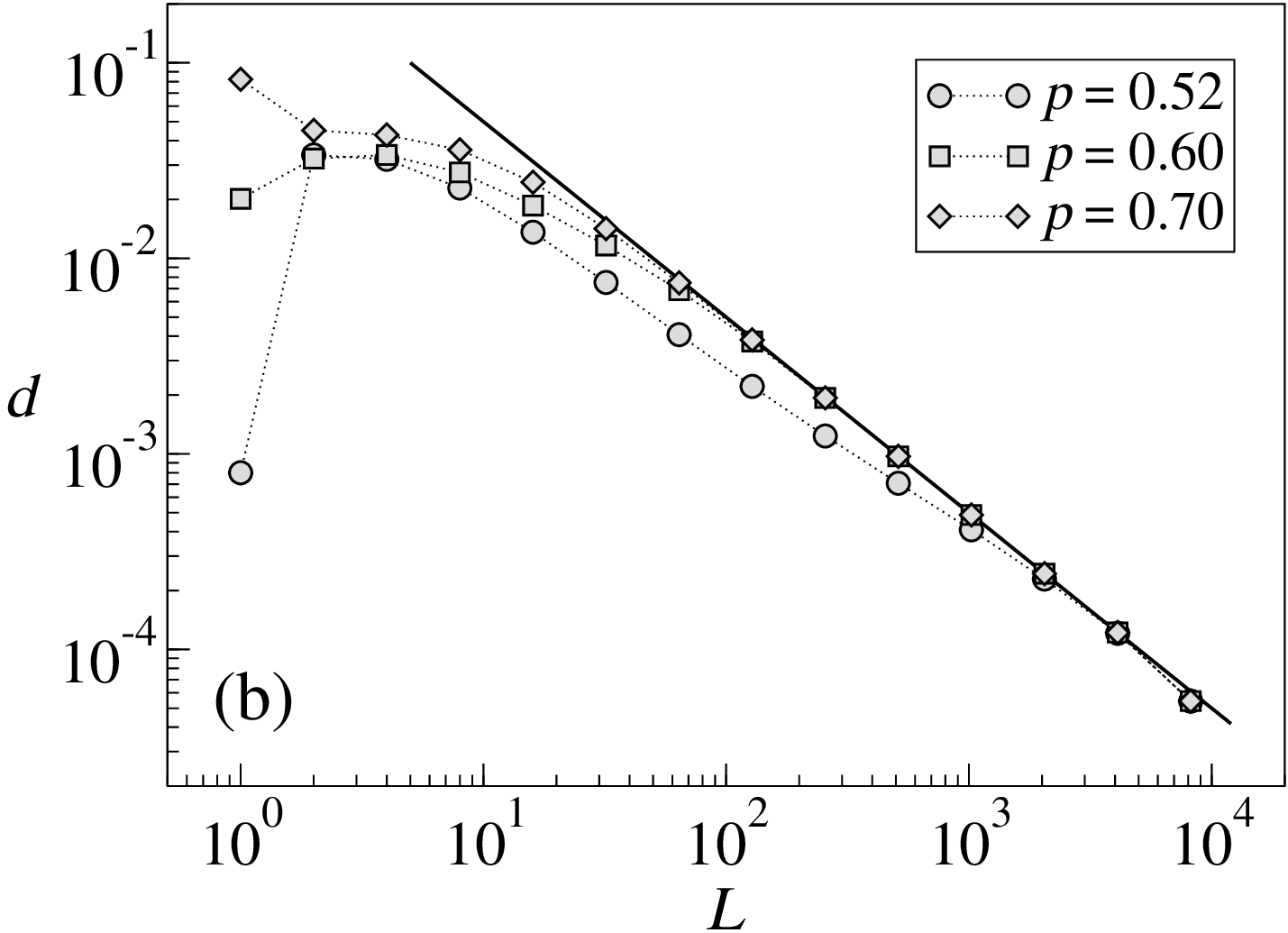}
\caption{a) The ratio $\xi:=\cR(p,L)/\cR\K(p)$ as a function of
the memory length $L$. b) To check the accuracy of
Eq.~\req{G-approx}, the difference $d:=G\K(p)-G(p,L)$ is shown
as a~function of $L$. Numerical results are shown as symbols,
dotted lines are used to guide the eye. Eq.~\req{G-approx}
suggests $d=1/(2L)$ which is shown as a solid line.}
\label{fig-memory}
\end{figure}

It is instructive to compute the exponential growth rate
$G(p,L)$ of an investor with the memory length $L$. If $p$
is fixed during the game, we have
\begin{equation}
\label{G-memory}
G(p,L)=\sum_{w=0}^L P(w\vert p,L)\,
\big[p\ln(1+f^*(w,L))+(1-p)\ln(1-f^*(w,L))\big].
\end{equation}
The compounded return is consequently $\cR(p,L)=\exp[G(p,L)]-1$.
This result can be compared with $\cR\K(p)$ of an investor with
the perfect knowledge of $p$ given by Eq.~\req{cR-Kelly}. In
Fig.~\ref{fig-memory}a, the ratio $\xi:=\cR(p,L)/\cR\K(p)$ is
shown as a function of $L$ for various $p$. As $L$ increases,
the investor's information about $p$ improves and $\xi\to1$.
Notice that when $p$ is smaller than a certain threshold (which
we numerically found to be approximately $0.63$), $\cR(p,L)<0$
for some $L$.

When $p$ is small, a very long memory is needed to
make a profitable investment: \emph{e.g.} for $p=0.51$,
$L\geq1\,761$ and for $p=0.52$, $L\geq438$ is needed. This
agrees with the experience of finance practitioners---according
to them the Kelly portfolio is sensitive to a wrong examination
of the investment profitability (for a scientific analysis of
this problem see~\cite{Sl99}). However, one should not forget
about the prior distribution $\pi(p)$ which is an efficient tool
to control the investment. For example, to avoid big losses in
a~weakly profitable game, $\pi(p)$ constrained to the range
$[0,3/4]$ can be used. In turn, if the game happens to be highly
profitable and $p>3/4$, such a choice of $\pi(p)$ reduces the
profit.

Due to its complicated form, $G(p,L)$ given by
Eq.~\req{G-memory} cannot be evaluated analytically. An
approximate solution can be obtained by replacing summation by
integration and noticing that for large values of $L$,
$P(w;p,L)$ is approaches the normal distribution
$g(w;\mu,\sigma)$ with the mean $\mu=Lp$ and the variance
$\sigma^2=Lp(1-p)$. Then we have
$$
G(p,L)\approx\int_{L/2}^L g(w;\mu,\sigma)h(w)\,\mathrm{d}w
$$
where $h(w)=p\ln\big[(2w+2)/(L+2)\big]+
(1-p)\ln\big[(2L+2-2w)/(L+2)\big]$. When $\sigma$ is small
compared to the length of the range where $h(w)$ is positive, we
can use the approximation
\begin{equation}
\label{approximation}
G(p,L)\approx h(\mu)+\frac{\sigma^2}{2}
\left.\frac{\partial^2 h}{\partial w^2}\right\rvert_{w=\mu}
\end{equation}
which is based on expanding $h(w)$ in a Taylor series;
a~detailed discussion of this approximation can be found
in~\cite{our}. It converges well when $h(w)$ is positive (and
thus bounded) in the range where $g(w;\mu,\sigma)$ differs from
zero substantially. Using $3\sigma$-range we obtain the
condition $L\gtrsim 9p(1-p)/(p-1/2)^2$; when the equality holds,
the final result obtained below has the relative error less than
10\%. Using $h(w)$ written above, after neglecting terms of
smaller order we obtain
\begin{equation}
\label{G-approx}
G(p,L)\approx G\K(p)-\frac1{2L}.
\end{equation}
In Fig.~\ref{fig-memory}, the quantity $d:=G\K(p)-G(p,L)$
computed numerically is shown as a function of $L$ for various
values of $p$, $d=1/(2L)$ following from Eq.~\req{G-approx} is
shown as a solid line. It can be seen that in its range of
applicability, the approximation works well. In addition,
Eq.~\req{G-approx} yields a rough estimate of the minimum memory
length needed to obtain $G(p,L)>0$ in the form
$L_{\min}\approx 1/[2G\K(p)]$.

When negative investment fractions are allowed,
$f^*=(2w-L)/(L+2)$ also for $w<L/2$. As shown in Appendix, one
can then obtain the series expansion of $G_{-}(p,L)$ (the
subscript denotes that $f<0$ is allowed) in powers of $1/L$
\begin{equation}
\label{G-better}
G_{-}(p,L)=G\K(p)-\frac1{2L}+\frac{11p(1-p)-1}{12p(1-p)L^2}-
\frac{20p^2(1-p)^2-5p(1-p)-1}{12p^2(1-p)^2L^3}+\dots.
\end{equation}
Notice that up to order $1/L$ it is identical with
Eq.~\req{G-approx}. When the terms shown above are used,
Eq.~\req{G-better} is highly accurate already for $L=20$.
Despite it is based on a different assumption, it can be also
used to approximate $G(p,L)$ discussed above; for small $L$ it
gives better results than the rough approximation
Eq.~\req{G-approx}.

\subsection{Another interpretation of the finite memory problem}
The optimal investment of a gambler with the memory length $L$
can be inferred also by the direct maximization of the
exponential growth rate. Then in addition to Eq.~\req{G-memory},
from the investor's point of view $G$ needs to be averaged over
all possible values of $p$, leading to
\begin{equation}
G(L)=\int_0^1\pi(p)\,\mathrm{d} p\sum_{w=0}^L P(w;p,L)\,
\big(p\ln[1+f(w,L)]+(1-p)\ln[1-f(w,L)]\big).
\end{equation}
This quantity can be maximized with respect to the investment
fractions $f(w,L)$ which is equivalent to the set of equations
$\partial G(L)/\partial f(w,L)=0$ ($w=0,\dots,L$). For
$\pi(p)=1$ in the range $[0,1]$ this set can be solved
analytically and yields the same optimal investment fractions
as given in Eq.~\req{f-memory}.

The statistical models described above use $\pi(p)$ as a model
for the gambler's prior knowledge of the winning probability
$p$. This knowledge can be caused by the lack of gambler's
information but also it can stem from the fact that $p$ changes
in time. Then $\pi(p)$ represents the probability that at
a~given turn, the winning probability is equal to $p$. Since
such an evolution of game properties is likely to occur in real
life, we investigate it in detail in the following paragraph.
We remind that the possible changes of $p$ in time are the key
reason why a gambler should use only a limited recent history of
the game.

\begin{figure}
\centering
\includegraphics[scale=0.26]{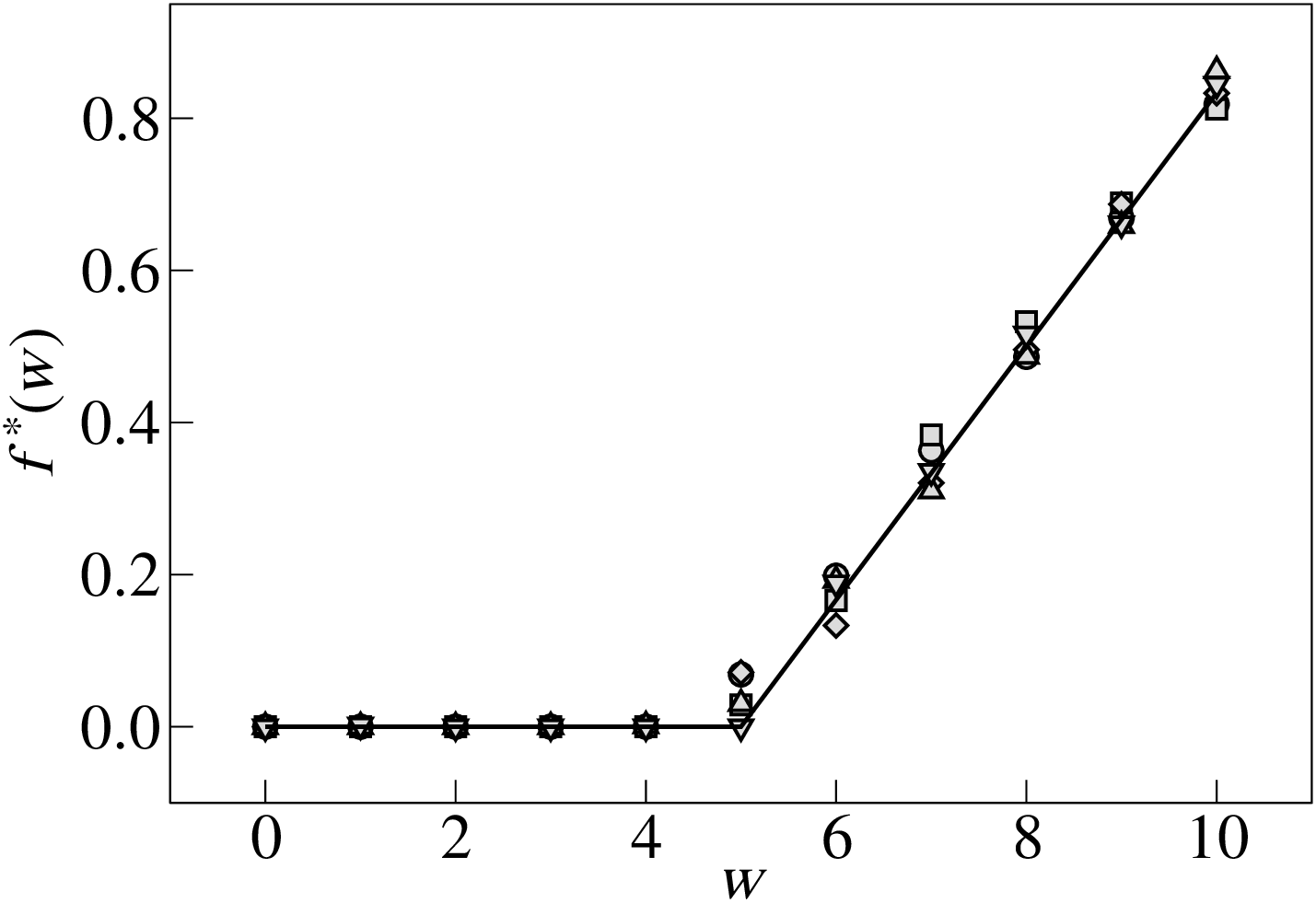}
\caption{A comparison of the analytical fractions $f^*(w,L)$
with results numerical maximization of $G$ for $L=10$ and
$N=1\,000\,000$. Various symbols are shown for five independent
realizations of the game.}
\label{fig-num-optimiz}
\end{figure}

As explained above, the evolution of $p$ can be incorporated in
$\pi(p)$. Consequently, if the changes of $p$ are slow enough to
assume that within time window of the length $L$ the winning
probability is approximately constant, the analytical results
obtained in the previous section still hold and the optimal
investment is given by Eq.~\req{f-memory}. To test this
conclusion, we maximized $G$ numerically with $f(w,L)$ as
variables ($w=0,\dots,L$) for five separate realizations of the
game, each with the length 1\,000\,000 turns and $L=10$. In each
realization, the winning probability changed regularly and
followed the succession $0.5\to1\to0\to0.5$. As a maximization
method we used simulated annealing~\cite{KGV83,Cer85}.
In~Fig.~\ref{fig-num-optimiz}, the result is shown together with
$f^*(w,L)$ given by Eq.~\req{f-memory} and a good agreement can
be seen. Thus we can conclude that with a proper choice of
$\pi(p)$, the analyzed model describes also a risky game with a
slowly changing winning probability.

\section{Conclusion}
In this work we examined maximization of the exponential growth
rate, originally proposed by Kelly, in various scenarios. Our
main goal was to explore the effects of diversification and
information on investment performance. To ease the computation,
instead of working with real assets we investigated simple risky
games with binary outcomes: win or loss. This allowed us to
obtain analytical results in various model situations.

First, in the case when multiple independent investment
opportunities are simultaneously available we proposed two
complementary approximations which yield analytical results for
the optimal investment fractions. Based on these results, we
proposed a simple framework to investigate the competition of
an uninformed investor (the outsider) who diversifies his
portfolio and an informed investor (the insider) who focuses on
one investment opportunity. We found the conditions when gains
from the diversification exceed gains from the additional
information and thus the outsider outperforms the insider.

Finally we investigated the performance of the Kelly strategy
when the return distribution (in our case the winning
probability of a risky game) is not known a priori. When the
past game outcomes represent the only source of information, we
found a simple analytical formula for the optimal investment. We
showed that for a weakly profitable game, a very long history is
needed to allow a profitable investment. As game properties may
change in time and thus the estimates obtained using long
histories may be biased, this is an important limitation. With
short period estimates suffering from uncertainty and long
period estimates suffering from non-stationarity, the Kelly
strategy may be unable to yield a profitable investment.

\section*{Acknowledgment}
This work was supported by Swiss National Science Foundation
(project 205120-113842) and in part by The International
Scientific Cooperation and Communication Project of Sichuan
Province in China (Grant Number 2008HH0014), Yu.~M.~Pis'mak was
supported in part by the Russian Foundation for Basic Research
(project 07-01-00692) and by the Swiss National Science
Foundation (project PIOI2-1189933). We appreciate interesting
discussions about our work with Damien Challet as well as
comments from Tao Zhou, Jian-Guo Liu, Joe Wakeling, and our
anonymous reviewers.

\appendix
\section{Developing a series expansion for $G_{-}(p,L)$}
Since for the exponential growth rate $G_{-}(p,L)$ there is no
closed analytical solution, we aim to obtain an approximate
series expression. As in Sec.~\ref{sec-memory}, $G_{-}(p,L)$ is
given by the formula Eq.~\req{G-memory} with
$f^*(w,L)=(2w-L)/(L+2)$ for $w=0,\dots,L$ and it can be
rearranged to
\begin{equation*}
G_{-}(p,L)=\ln\frac2{L+2}+
\sum_{w=0}^L\frac{L!\,p^w(1-p)^{L-w}}{w!(L-w)!}
\big[p\ln(w+1)+(1-p)\ln(L+1-w)\big],
\end{equation*}
To get rid of the logarithm terms we use
$\ln z=-z\int_0^{\infty} (\gamma+\ln t)\mathrm{e}^{-tz}\dd t$
where $\gamma\approx0.577$ is the Euler's constant. This formula
follows directly from the usual integral representation of the
Gamma function. Consequently, by the substitution
$\tau:=\mathrm{e}^{-t}$ we obtain
\begin{equation}
\label{identity}
\ln z=-z\int_0^1\big[\gamma+\ln(-\ln\tau)\big]\tau^{z-1}\dd\tau.
\end{equation}
After exchanging the order of the summation and the integration
in $G_{-}(p,L)$ it is now possible to sum over $w$, leading to
$G_{-}(p,L)=\ln\big[2/(L+2)\big]+R(p)+R(1-p)$ where
\begin{eqnarray*}
\label{R}
R(p)&=&\int_0^1\big[\gamma+\ln(-\ln\tau)\big]
T(\tau,p,L)\,\dd\tau,\\
T(\tau,p,L)&=&p(\tau p+1-p)^{L-1}\big[1-p+\tau p(1+L)\big].
\end{eqnarray*}
The substitution $\tau:=1-\varrho/L$ allows us to obtain
series expansions of $T(\tau,p,L)$ and $\ln(-\ln\tau)$ in powers
of $1/L$ which after integration lead to Eq.~\req{G-better}.

\end{document}